\documentstyle[11pt]{article}
\begin{document}
\baselineskip=22pt
\pagestyle{plain}
\thispagestyle{empty}
\null
\normalsize
\def\ref{\noindent\hangindent=3em}
\def\arcsec{\prime\prime}
\def\kms{\mbox{$\rm km~s^{-1}$}}

{\Large
\centerline{\bf Giant Low Surface Brightness Galaxies~:~Evolution in Isolation}
}

\vspace{0.15cm}

\centerline{Mousumi Das}

\centerline{Indian Institute of Astrophysics, Bangalore.}

\vspace{3mm}

\centerline{Abstract}

Giant Low Surface Brightness (GLSB) galaxies are amongst
the most massive spiral galaxies that we know of in our Universe.
Although they fall in the class of late type spiral galaxies, their properties
are far more extreme. They have very faint stellar disks that are extremely
rich in neutral hydrogen gas but low in star formation and hence low in
surface brightness. They often have bright bulges that are similar to those
found in early type galaxies. The bulges can host low luminosity Active
Galactic Nuclei (AGN) that have relatively low mass black holes. GLSB
galaxies are usually isolated systems and are rarely found to be interacting
with other galaxies. In fact many GLSB galaxies are found under dense
regions close to the edges of voids. These galaxies have very massive dark
matter halos that also contribute to their stability and lack of evolution. In
this paper we briefly review the properties of this unique class of galaxies
and conclude that both their isolation and their massive dark matter halos
have led to the low star formation rates and the slower rate of evolution in
these galaxies.

\noindent keywords : Galaxies: evolution—galaxies: nuclei—galaxies: active—
galaxies: ISM—galaxies: spiral—cosmology: dark matter.
 
\section {Introduction}

Giant low surface brightness (GLSB) galaxies are some of the largest spiral galaxies in 
our nearby universe. However, for decades these galaxies remained undetected in galaxy surveys. This is 
because their optically dim stellar disks have a brightness that lies close to or 
below the brightness of 
the night sky. Early photometric studies of galaxies determined that the central surface brightness of galaxies 
always lies
above the brightness limit of $\mu_{B}(0)~\sim~21.65~mag~arcsec^{-1}$; this is called the Freeman's
Law (Freeman 1970). However, Disney (1976) predicted that there may be a population of galaxies that lie below this
brightness limit. His study indicated that the brightness of the night sky biased observations against detecting 
low luminosity galaxies such as low surface brightness (LSB) galaxies. 
Decades later with better telescopes, it became possible to detect sources well below the Freeman limit.
Recent surveys have revealed that LSB galaxies contribute a very significant fraction of the total 
galaxy number density in the local (z~<~0.1) universe (Trachternach et al. 2006). 
The first large LSB or GLSB galaxy to be detected was Malin~1. It was accidentally discovered in a survey of
low luminosity galaxies (Bothun et al. 1987). In the following years several GLSB galaxies 
were discovered (Bothun et al. 1990; Schombert et al. 1992; Sprayberry et al. 1993). 

Although morphologically LSB galaxies span a very wide range from dwarfs and irregulars to very large disk 
galaxies (McGaugh et al. 1995),  they are broadly of two
types (1)~LSB dwarf and irregular galaxies and (2)~disk LSB galaxies of which the larger ones are called GSLB galaxies. 
Regardless of their size or morphology,
all LSB galaxies share the following common characteristics; poor star formation rates, low metallicities, diffuse 
stellar disks and extended HI gas disks (Impey \& Bothun 1997). The LSB dwarfs 
and irregulars form the larger fraction of LSB galaxies; studies indicate that they are the most dominant component
of the faint end of the luminosity function in our local Universe (Geller et al. 2012). Although optically they 
are often difficult to detect, they are easily detected in HI surveys. Unlike the more isolated larger LSB galaxies, dwarf
LSB galaxies can be found in relatively densely populated parts of the universe such as galaxy
groups (Sabatini, Davies, Scaramella et al. 2003). The fainter ones are generally found in underdense environments such 
as nearby voids (e.g. Pustilnik et al. 2011). 

Disk LSB galaxies are not as common as LSB dwarf galaxies. Studies show that they can span a range of sizes 
(e.g.~Beijersbergen et al. 1999) but their Hubble morphological type is well defined as late type spirals - Sc or Sd. 
In some cases, where there is a very prominent bulge (e.g.~UGC~6614, (R)SA(r)) the galaxies are classified as 
early type spirals (see Figure~1a). But their low surface brightness disks and other properties such as extended HI
disks, confirm that they are extreme late type spirals. The really large LSB galaxies are usually referred 
to as giant LSB (GLSB) galaxies and are generally isolated systems (Bothun et al. 1993). They are found to lie 
closer to the walls of voids (Rosenbaum et al. 2009). This review will focus only on GLSB galaxies;
their nuclear and disk properties are very distinct from the LSB dwarfs galaxies or smaller LSB disk galaxies. 

Apart from their rather exotic and rare nature, GLSB galaxies provide an interesting sample to study
how massive dark matter halos can affect star formation and even nuclear activity in galaxies. They also provide
an opportunity to understand how galaxies evolve in relative isolation. In this paper, we will review the overall
properties of GLSB galaxies and then discuss how their dark matter content and isolation shapes their evolution.

\section {Structural Properties~:~Disks and Bulges}

GLSB galaxies, like high surface brightness (HSB) spiral galaxies, have radial surface brightness profiles that follows
an exponential form i.e. $\Sigma(r)=\Sigma_{0}e^{-r/\alpha}$ where r is the disk radius and $\alpha$ is the disk 
scale length (de Blok, van der Hulst \& Bothun 1995). The main difference is that the extrapolated disk central 
brightness $\mu_{B0}$ lies between 22 to 23~mag/arcsec$^2$, which is much lower thani thant observed in regular galaxies 
and lower than the the Freeman limit. Some GLSB galaxies have prominent bulges 
(e.g.~UGC~6614 and UGC~9024; McGaugh and Bothun 1994). In such cases the surface brightness profile can
be fitted with two exponential profiles. The stellar disks are usually very large and 
greater than $\alpha\sim$10~Kpc (McGaugh and Bothun 1994); but the smaller LSB disk galaxies have stellar disks with
$\alpha\sim$4-5~Kpc.

The low luminosity of the LSB disks indicates a low surface density of stars. The maximum disk 
hypothesis can be applied to these stellar disks to determine upper limits to the mass to luminosity (M/L) 
ratios and stellar disk masses. These upper mass limits show that LSB galaxies have lower stellar surface mass densities 
than normal
spiral galaxies (de Blok, McGaugh \& Rubin 2001). Mass modeling of LSB galaxy rotation curves using stellar population 
synthesis models also indicates a high M/L and low stellar disk surface densities (Swaters et al. 2000). The diffuse disks 
of GLSB galaxies contrasts sharply with their bright bulges (Pickering et al. 1999). A good example 
is UGC~6614 (Figure~1a); its LSB stellar disk is barely visible except for the tightly wound spiral arms. GLSB galaxy disks 
often have spiral arms that can be followed well into their disks ( e.g~NGC~5905, Figure~1b). But the arms are notably
"thin" indicating that they are not associated with very massive star forming regions as seen in HSB galaxies. In 
some galaxies the arms are practically absent (e.g. UGC~1922; O'Neil \& Schinnerer 2003).  Bars are not common
in GLSB galaxies and only 10-15\% of LSB disk galaxies have bars in the centers of their disks. Bars and evidence for 
tidal interactions are weak in LSB galaxies (Mihos et al. 1999). 
The overall lack of strong disk instabilities suggests that the dark halo must be strong in all LSB galaxies and 
especially for the more massive GLSB galaxies (Mayer \& Wadsley 2004). 

Although most LSB disk galaxies appear to be late type systems with relatively small bulges, their near-infrared images 
clearly show that a large fraction of them have significant bulges (Galaz et al. 2002; Beijersbergen et al. 1999). These
bulges are very prominent in the GLSB galaxies (e.g. UGC~6614, Malin~1, Malin~2) and are usually classical bulges. In 
some GLSB galaxies such as UGC~2936, the bulge is boxy and is possibly a pseudobulge. UGC~2936 also has 
significant star formation in its disk; both features indicate that there maybe ongoing secular evolution in this 
galaxy resulting in a more oval than spherical bulge (Kormendy \& Kennicutt 2004). GLSB galaxy bulges are notably bright 
and studies suggest that they are similar to the bulges of HSB galaxies both in stellar population and metallicities
(Morelli et al. 2012). Thus their bulges are suprisingly very different from their low luminosity disks (McGaugh et al. 1995)
both in structure and composition. This suggests that the disks and bulges have different evolutionary histories and did 
not co-evolve.
 
\section {Gas Content~:~HI and Molecular Gas}

One of the main features of all LSB galaxies (dwarf, irregulars and disks) is that they have large amounts of HI gas
(O'Neil et al. 2004). The HI masses are so large in GLSB galaxies ($10^{9} - 10^{10}$) that they are comparable to the typical 
stellar masses of late type galaxies (Matthews et al. 2001) and make up a considerable fraction of their baryonic mass
(McGaugh et al. 2000).  The HI profiles and maps show that the gas is far more extended than the stellar disks 
(Pickering et al. 1997; de Blok, McGaugh, van der Hulst van 1995).  
But although the HI masses are large, the HI gas surface densities $\Sigma$ in $M_{\odot}~pc^{-2}$ are much lower than that
of normal late type spirals (de Blok, McGaugh \& van der Hulst 1996). The low $\Sigma$ of the HI distribution has
important implications for the star formation properties of GLSB and LSB galaxies in general (see next section). The HI 
disks are often flared or warped (Matthews \& Kenneth 2003; Pickering et al. 1999) or lopsided (Das et al. 2007). Such 
features are more commonly found at larger radii where the stellar disk surface density decreases and the HI disk begins to 
be more dominant. Thus    
they appear more prominent in the outer disks of spirals where the effect of the halo becomes more prominent (Reichard et al. 2008).
The HI rotation curves of GLSB galaxies are usually slowly rising but flatten at relatively high velocities 
or in some cases continue rising well into their disks. This is a clear indication of their large dark matter content
which is also seen in their relatively high mass to luminsoity (M/L) ratios. 

Molecular hydrogen gas ($H_2$) is rare in LSB galaxies (de Blok \& van der Hulst 1998; Braine, Herpin \& Radford 2000)
but is this is not suprising considering their low star formation rates and low metallicities (Schombert et al. 1992). 
$H_2$ has been detected in only a handful of GLSB galaxies and so far never detected in LSB dwarf galaxies (O’Neil et al.
2000; Matthews \& Gao 2001; O’Neil et al. 2003; Matthews et al.2005; Das et al. 2006). The masses detected are in the 
range $10^{8} - 10^{9}~M_{\odot}$ which though significant, are low compared to the galaxy HI gas masses and also low compared 
to the M($H_2$)/M(HI) fractions observed in normal galaxies. The molecular gas distribution has been studied in very few
galaxies (Malin~2, Das et al. 2010; UGC~1922, O'Neil \& Schinnerer 2003).  In UGC~1922 it is mainly concentrated within the inner disk 
but in Malin~2 it is more extended and supports star formation (Pickering et al. 1997). In all these studies CO(1--0) 
or CO(2--1) was used as a tracer for molecular hydrogen. It is possible that the $H_2$ exists as cold molecular gas 
or "dark gas" and resides within photodissociation regions, where the CO molecule has been dissociated by ionizing stellar 
radiation but $H_2$ molecules still exist (Planck Collaboration et al. 2011; Grenier et al. 2005). In such cases the cold molecular
hydrogen gas may exist as diffuse, dark gas but will go undetected in millimeter CO observations. But this is unlikely to
be the case in GLSB or dwarf LSB galaxies as these galaxies are so poor in dust and low in metallicities that the neutral gas
will not be able to cool enough or fast enough to form $H_2$ molecules via dust particles adsorption.   Thus,
the lack of $H_2$ gas in these galaxies is probably real (Gerritsen \& de Blok 1999). 

\section {Star formation Properties}

In general LSB galaxies are low in star formation (O'Neil, Oey \& Bothun 2007) especially when measured per unit stellar 
mass (Schombert, Maciel, McGaugh 2011). This is especially true for LSB dwarf galaxies which are some of the most
dark matter dominated systems in our Universe. H$\alpha$ observations of the larger LSB disk galaxies and epsecially GLSB
galaxies reveal that they have patches of localized star formation (e.g. Auld et al. 2006)
associated with faint spiral arms or distributed sporadically over their disks. Suprisingly, in spite of their low star 
formation rates (SFR) many studies clearly show that 
their colors i.e. (B-V) or (B-R), are ususally blue and comparable to normal galaxies (van der Hulst et al. 1993). This is partly 
due to the larger number of localized pockets of star formation in GLSB disks, which are generally large. But it may also be
due to the fact that the V or R magnitude of these galaxies are low since their stellar disks are diffuse.
So the color (i.e. (B-V) or (B-R)) is high even though the star formation rates are low. A significant population of 
red LSB galaxies have also been 
observed. But these are generally smaller galaxies and do not have GLSB disks (O'Neil, Bothun, Schombert 2000).  

The star formation in GLSB galaxies is predominantly associated with their bulges and very inner disks. In contrast to the
LSB disks, the bulges have a younger population of stars and show ongoing star formation (Morelli et al. 2012). The metal 
content is often solar and similar to HSB galaxies. Evidence for bulge star formation is also seen in the X-ray  
observations of GLSB galaxies as the diffuse emission associated with star formation is mainly confined to 
their bulges (Das et al. 2009). Thus in general GLSB bulges are similar to the bulges of HSB galaxies in terms of star formation 
properties. In some galaxies, such as the very massive and distant GLSB galaxy Malin~1, there is both a bright bulge and inner disk 
that appears to be similar to normal galaxies in surface brightness and structure (Barth 2009). Galex UV observations of the disks
indicate that GLSB galaxies have UV emission that extended well beyond their optical disks (Wyder \& Treyer 2011). The extended
morphology and UV colors indicate that star formation occurred in bursts and had an efficiency far lower than that found in 
normal galaxies (Boissier et al. 2008). Thus the bright bulges and LSB disks of GLSB galaxies appear to have had very 
different star formation histories. 

Several studies have tried to understand the lack of star formation in LSB galaxies. In general the 
presence of copious amounts of neutral hydrogen gas in galaxies indicates that they have the capacity to form stars through 
cloud collapse and star formation. But although LSB galaxies are gas rich, they are very poor in star formation. Both their
star formation rates (SFRs) and metallicities are low. Also, as mentioned in the previous section molecular gas, which is always 
found associated with star formation, is rarely detected in these galaxies.  
One of the reasons for the low SFR is the lack of strong global instabilities as well as local disk instabilities. 
Both types of instabilities induce cloud collisions and shocks that result in gas compression and gas cooling; these processes
can trigger star formation. The global disk    
instability criterion $X_{m}$ depends on $\Sigma$ (see Mihos, de Blok \& McGaugh 1997). If $\Sigma$ is low, 
global nonaxisymmetric disk instabilities such as spiral arms and bars cannot form. In GLSB galaxies, the low surface densities 
of both the stellar and HI disks results in the formation of only small bars and weak spiral features. Such weak nonaxisymmetric instabilities 
cannot drive the large scale massive star formation that is commonly observed in HSB galaxies. Alternatively, localized
star formation can be produced by local disk instabilities. The latter is measured by Toomre’s Q parameter, which also depends 
on the $\Sigma$ of the HI gas disk. But again since $\Sigma$ is low in GLSB galaxies, local instabilities are also difficult to form
(van der Hulst et al. 1993; Das et al. 2010). Localized star is observed in some GLSB disks; it may be due to distant tidal interactions
(e.g. NGC~5905; van Moorsel 1982) or small accretion events (e.g. Malin~2; Pickering et al. 1997).   
The other two factors responsible for low SFRs in LSB galaxies are their low metal content and low dust masses (Gerritsen \& de Blok 1999).  
Metals are important for gas cooling and the presence of dust is important for the formation of $H_2$ molecules. The lack of both,
which is intimately related to the low SFRs in GLSB galaxies, contributes to the slow evolution of these galaxies.

\section {Nuclear Properties}

LSB dwarfs, irregulars and smaller disk galaxies show very little nuclear activity. But the larger LSB disk galaxies often 
show nuclear star formation and some have Active Galactic Nuclei (AGN). A large fraction of GLSB galaxies are relatively bulgeless
(McGaugh \& Bothun 1994) but there is often a bright core due to nuclear star formation. In most bulgeless galaxies, the
nuclear activity appears as a luminous point in an often featureless low luminosity disk (Matthews et al. 1999). Strong
H$\alpha$ and [OI] emission lines in their spectra indicate ongoing nuclear star formation. High resolution ground based imaging
indicates that the star formation is often in the form of  ~Kpc scale
rings, as seen in the GLSB galaxy NGC~5905 (Comeron et al. 2010). Hubble space telescope (HST) observations have shown that the 
nuclear star formation can lead to the formation of compact nuclear star clusters that may sometimes co-exist with AGN activity
(Seth et al. 2008). This activity can 
contribute to the formation of a central massive object (CMO) and lead to the 
buildup of a bulge in an otherwise bulgeless galaxy (Davies, Miller \& Bellovary 2011).

A significant fraction of bulge dominated GLSB galaxies show AGN activity (Sprayberry et al. 1995; Galaz et al. 2011). 
This is not suprising as studies indicate that the growth of nuclear black holes in galaxies is intimately linked to the 
growth of their bulges (e.g. Silk \& Rees 1998; Heckman et al. 2004). The strong correlation of
black hole mass (M) with bulge mass or bulge luminosities in galaxies ($M - \sigma$) is due to this supermassive
black hole (SMBH) - bugle co-evolution (e.g. Gultekin et al. 2010). Early studies by Schombert (1998) suggested that half of all
GLSB galaxies have AGN but later studies of larger samples show that only about 15\% have AGN (Burkholder, Impey, Sprayberry 2001).
AGN activity will give rise to relatively broad H$\alpha$ and H$\beta$ emission lines that can be detected in the optical spectra
of the galaxies, whereas star formation will result in narrower lines emission (Schombert 1998). Depending on whether the lines 
have a broad component or not, the nuclei are either Seyfert~1 (Sy~1) or Seyfert~2 (Sy~2) type AGN.
We examined at the SDSS spectra of a large sample of LSB disk galaxies collected from the literature and conclude that the 
fraction of GLSB galaxies hosting AGN activity is not more than 10\% (Subramanium et al. 2013, in preparation). The presence
of an AGN can also be confirmed from X-ray observations. Compact X-ray emission associated with the nucleus of a galaxy is a
signature of AGN activity and has been detected in a few GLSB galaxies such as UGC~6614 (Naik et al. 2009), UGC~2936 
(Das et al. 2009a) and UGC~1455 (Das et al. 2009a). Radio emission associated with AGN activity has also been detected in several
GLSB galaxies (Das et al. 2007; Mishra et al. 2013 in preparation). For example, in UGC~6614 the nuclear emission has a compact
core at millimeter wavelengths (110~GHz, Das et al. 2006) as well as longer wavelengths 
(1.4~GHz, Das et al. 2009b, also see Figure~1a). The spectral
index $\alpha$ between these frequencies is close to 0 (i.e. a flat core) suggesting that the source is definitely an AGN. At
lower frequency of 610~MHz, the emission is extended beyond the core (Das et al. 2009b) and resembles jets or outflows associated with the 
AGN activity (Mishra et al. 2013, in preparation). Extended emission is also seen in 2327-0244 (or UM~163) and resembles a one sided jet
(Das et al. 2009b). Another example of compact radio emission associated with an AGN is Malin~2 (see Figure~1b).

Not much is known about the black hole (BH) masses in LSB galaxies. The large bulges of GLSB galaxies suggests that they may harbour
SMBH's (i.e. BH's with masses larger than $10^{6} - 10^{7}~M_{\odot}$) (Rees 1984). The bulge mass and galaxy sizes suggest that 
the BH masses may be comparable to the SMBHs in ellipticals.  However, studies of the optical spectra of Sy~1 type nuclei in GLSB
galaxies suggest that the BH masses lie in the $10^{5} - 10^{6}~M_{\odot}$ range instead (Ramya et al. 2010; also see Table~2). 
Sy~1 AGN ususally have a broad 
component in their H$\alpha$ or H$\beta$ emission lines due to the effect of the broad line region (BLR) clouds on the emission 
from the AGN. Assuming virial equilibrium in the BLR region, the BH mass can be estimated from the width of the broad component of 
the H$\alpha$ or H$\beta$ emission lines and the line luminosities (Greene \& Ho 2007). This is a virial approximation and hence does 
not give as accurate a BH mass as reverberation mapping or stellar kinematic techniques. But it does give an estimate of the BH mass 
range. The studies done so far (Table~2) indicate that the the masses lie more in the intermediate black hole (IMBH) mass range rather
than SMBH rangei (Ramya et al. 2011). 
When the bulge velocity dispersions are plotted against the BH masses to obtain the M-$\sigma$ correlation for these galaxies, most 
GLSB galaxies are found to lie below the standard M-$\sigma$ correlation (Gultekin et al. 2011). The points have a scatter but their 
distribution suggests that as the BH masses aproach $10^{7}~M_{\odot}$, they begin to follow the M-$\sigma$ correlation for bright galaxies.
Thus only GLSB galaxies that have SMBHs of mass larger than $10^{7}~M_{\odot}$ follow the same evolutionary track as normal galaxies 
on the M-$\sigma$ relation (Subramanium et al. 2013, in preparation).   

\section {Dark Matter in GLSB galaxies}

LSB galaxies are one of the most dark matter dominated systems in our nearby Universe (de Blok \& McGaugh 1997). The only class of 
galaxies that have comparable 
baryon to dark matter ratios are dwarf spheroidal galaxies, which are the prime targets for dark matter particle searches 
(e.g. Abdo et al 2010). Traditionally low luminocity galaxies have been associated with massive dark matter halos 
(Kormendy 1990; Kormendy \& Freeman 2004). In the case
of GLSB galaxies the main observational evidence that suggests the the presence of massive dark halos in these galaxies is the 
large M/L ratios derived from rotation curve fitting methods (McGaugh \& de Blok 1998).  Even when a maximum disk is 
applied to the galaxies, the dark matter is substantially more than the stellar and gas mass combined. The rotation curves in GLSB
galaxies are slowly rising and flatten at velocities of 200 - 300~km/s, indicating the presence of very massive halos that have 
large dynamical masses (Pickering et al. 1997). Another approach is to derive the M/L ratio from the galaxy colors (Bell et al. 2003).
This method suggests that like most faint galaxies, LSB galaxies lie at the lower end of the luminosity curve and their masses were
undersestimated in earlier photometric surveys.  

One of the suprising and puzzling aspects of dark matter in LSB galaxies is that the distribution does not follow predictions of cold 
dark matter (CDM) theories of structure formation. CDM predicts that the dark matter profiles of galaxies should have cuspy cores which 
follow the NFW profile(Navarro, Frenk, White 1996). However, rotation curve studies show that in general LSB galaxies have profiles 
that are closer to isothermal or flat core in shape; there is no evidence for cuspy halos in these galaxies 
(McGaugh, Rubin, \& de Blok 2001). Mass modeling of the hybrid H$\alpha$ and HI rotation curves indicate core sizes of the order of 1~Kpc 
(de Blok et al. 2001). However, beam smearing in rotation curve modeling has always cast some doubts against the isothermal core model , as it 
can result in inaccuracies in estimating velocities in the very inner regions. But further studies using two dimensional high resolution optical 
velocity fields of LSB galaxies confirmed that the the NFW model cannot work for the dark matter halos in low luminosity galaxies 
(Kuzio de Naray, McGaugh, de Blok 2008). 

\section {Formation of GLSB galaxies}

Not a lot is known about the formation of LSB galaxies and why they are so different from HSB galaxies. 
The early theoretical study by Hoffman et al. (1992), which
was based on the hierarchical picture of galaxy formation, showed that GLSB galaxies will form in voids due to
rare 3$\sigma$ perturbations in underdense regions of the universe. The smaller amplitude peaks (1-2$\sigma$) can
be associated with the smaller dwarf LSB galaxies (Dekel \& Silk 1986). Unlike GLSB galaxies their distribution does
not depend so much on the large scale structure and their distribution is more uniform. In this picture, the GLSB 
galaxies that form in underdense regions, will have normal 
bulges but extended, unevolved disks. Although observations do indicate that GLSB galaxies are often isolated, they do
not always populate the interior of voids (Bothun et al. 1986). Instead recent studies using the SDSS show that the larger 
LSB galaxies are in fact more clustered towards the outer edges of the void walls and filaments (Rosenbaum et al. 2009).
They  reside in relatively isolated environemnts compared to 
HSB galaxies (Ceccarelli et al. 2012) and their relative isolation must have contributed to their slow evolution.

Another possibility is that GLSB galaxies formed in halos with large angular momentum (Jimenez et al. 1998). However, 
in this picture GLSB galaxies form rather late at z=0.7. Other models
of protogalaxies with large angular momentum also form extended galaxy disks that are very similar to GLSB galaxies 
(Dalcanton et al. 1997). These results indicate that there must some connection between the halos of GLSB galaxies
and their properties, as well as evolution. But halo spin alone cannot account for their relative isolation and poor 
evolution. Instead it is possible that a combination of both isolation and halo
spin are important to explain the unique properties of these galaxies and how they fit into the picture of hierarchical 
structure formation.

\section {Evolution of GLSB galaxies}

Galaxies evolve by two methods. (i)~The first is through external triggers such as interactions with other galaxies 
and accretion of smaller galaxies. Even 
distant interactions can cause significant non-axisymmetric disk perturbations that result in cloud collisions and gas infall 
from the disks into the galaxy centers. Both distant interactions and mergers result in an overall increase in star formation.
Interactions can lead to the formation of global disk instabilities such as bars and spiral arms. These perturbations further
enhance the rate of gas infall to nucleus, often leading to nuclear starburst (Friedli \& Benz 1993). All these 
processes can result in the build-up of central mass concentrations that often weaken and even destroy bars in the centers
of galaxies (Norman, Sellwood, Hasan 1996; Das et al. 2003). These processes also contribute to the growth of supermassive 
black holes and AGN in galaxy nuclei. The GLSB galaxy NGC~5905 is a good example (Figure 1a); the galaxy is  
interacting with the relatively distant spiral galaxy NGC~5908 and the interaction has resulted in the formation of a strong bar.
The bar is
associated with a bright bulge that shows star formation and relatively tightly wound spiral arms. (ii)~The second mode of 
galaxy evolution is the slow secular form, which is due to internal processes in galaxies (Kormendy \& Kennicutt 2004) and is 
called the secular evolution of galaxies. In this process, bars play a very prominent role as they exert gravitational torques 
oni disk gas resulting in the pileup of gas in 
inner/outer rings and in the centers of galaxies disks. Bars can become unstable resulting in the formation of disky bulges,
called pseudobulges. The GLSB galaxy UGC~2936 is a good example of secular evolution in an isolated galaxy. It has ongoing 
disk star formation which is evident from the radio continuum emission maps (Das et al. 2009b) and a disky bulge (Pickering et al. 1999). 

The two most crucial factors that affect the evolution of GLSB galaxies is their isolation and their massive dark matter halos.
Since their discovery more than two decades ago, it has been clear that GLSB galaxies are relatively isolated compared to HSB 
galaxies (Sprayberry et al. 1995). More recent studies using the Sloan Digital Sky Survey (SDSS) confirm that GLSB galaxies are
generally isolated and often found close to the edges of voids (Galaz et al. 2011); Rosenbaum et al. 2009). The lack of interactions  
thus results in less disk instabilities and hence lower star formation rates in GLSB galaxies. This leads to a slower pace of 
evolution in these galaxies. But perhaps the most important
factor influencing GLSB galaxy evolution is the massive dark matter halos found in these systems. It has been known from the 
early work of Ostriker \& Peebles (1973) that galaxies become 
more stable against the formation of disk instabilities when there is a massive halo component. Later numerical studies have shown
that this result is very relevant to LSB galaxies (Mayer \& Wadsley). In fact simulations show that the dark matter halos 
stabilise disks even in tidal
interactions with other galaxies (Mihos et al. 1997). Thus the dark matter halos slow down the formation of global
and local non-axisymmetric instabilities and hence lead to a slower rate of galaxy evolution in GLSB galaxies.     

\section {Conclusions}

\noindent
{\bf 1.}~GLSB galaxies are an extreme form of late type spiral galaxies. They are characterised by very faint, extended disks
that are low in stellar density but very rich in neutral hydrogen gas. They have low metal content and low star formation rates.
Molecular gas, which is associated with star formation, is rarely detected in these galaxies. 

\noindent
{\bf 2.}~Their disks are embedded in very massive dark matter halos that make them very stable against disk instabilities such
as bars and spiral arms. Hence star formation is low;  this leads to very low surface brightness disks that often lie close to 
or below the brightness of the night sky. 

\noindent
{\bf 3.}~Their nuclei sometimes host AGN activity. Indirect estimates of the associated black masses yield values 
$10^{5} - 10^{7}~M_{\odot}$. The AGN can also be detected in X-ray and radio emission. 

\noindent
{\bf 4.}~The distribution, morphology and disk dynamics of GLSB galaxies indicate that both their isolation and dominant 
dark matter halos 
contribute to their extremely slow rate of evolutuion compared to normal galaxies.    

\vspace{3mm}

\noindent
{\bf Acknowledgements :}\\
This work has used a mid-infrared image of NGC~5905 which was based on observations made with the Spitzer Space Telescope,  
which is operated by the Jet Propulsion Laboratory, California Institute of Technology under a contract with NASA. This research
has made use of the NASA/IPAC Extragalactic Database (NED) which is operated by the Jet Propulsion Laboratory, California Institute
of Technology, under contract with the National Aeronautics and Space Administration. This work has also made use of the
NRAO VLA NVSS maps of UGC~6614 and Malin~2. M.D. would like to thank Alice Quillen for the R band image of UGC~6614 and also 
Almuden Alonso Herrero for help with the  Spitzer image of NGC~5905.

\vspace{3mm}

\noindent
References :\\
Auld, R., de Blok, W. J. G., Bell, E., Davies, J. I. 2006, MNRAS, 366, 1475.\\
Barth, A. J. 2007, AJ, 133, 1085.\\
Beijersbergen, M., de Blok, W. J. G., van der Hulst, J. M. 1999, A\&A, 351, 903.\\
Bell, E. F., McIntosh, D. H., Katz, N., Weinberg, M. D. 2003, ApJS, 149, 289.\\
Boissier, S., Gil de Paz, A., Boselli, A., Buat, V. 2008, ApJ, 681, 244.\\
Bothun, G. D., Beers, T. C., Mould, J. R., Huchra, J. P. 1986, ApJ, 308, 510.\\
Bothun, G. D., Impey, C. D., Malin, D. F., Mould, J. R. 1987, AJ, 94, 23.\\
Bothun, G. D., Schombert, J. M., Impey, C. D., Schneider, S. E. 1990, ApJ, 360, 427.\\
Bothun, G. D., Schombert, J. M., Impey, C. D., Sprayberry, D., McGaugh, S. S. 1993, AJ, 106,
530.\\
Braine, J., Herpin, F., Radford, S. J. E. 2000, A\&A, 358, 494.\\
Burkholder, V., Impey, C., Sprayberry, D. 2001, AJ, 122, 2318.\\
Ceccarelli, L., Herrera-Camus, R., Lambas, D. G., Galaz, G., Padilla, N. D. 2012, MNRAS,
426L, 6.\\
Comerón, S., Knapen, J. H., Beckman, J. E., Laurikainen, E. et al. 2010, MNRAS 402, 2462.\\
Dalcanton, J., Spergel, D. N., Gunn, J. E., Schmidt, M., Schneider, D. P. 1997, AJ, 114,
2178.\\
Das, M., Boone, F., Viallefond, F. 2010, A\&A, 523A, 63.\\
Das, M., Reynolds, C. S., Vogel, S. N., McGaugh, S. S., Kantharia, N. G. 2009a, ApJ, 693,
1300.\\
Das, M., Kantharia, N. G., Vogel, S. N., McGaugh, S. S. 2009b, ASPC, 407, 167.\\
Das, M., Kantharia, N., Ramya, S., Prabhu, T. P., McGaugh, S. S., Vogel, S. N. 2007, MNRAS,
379, 11.\\
Das, M., O’Neil, K., Vogel, S. N., McGaugh, S. 2006, ApJ, 651, 853.\\
Das, M., Teuben, P. J., Vogel, S. N., Regan, M. W., Sheth, K., Harris, A. I., Jefferys, W. H.
2003, ApJ, 582, 190.\\
Davies, M. B., Miller, M. C., Bellovary, J. M. 2011, ApJ, 740L, 42.\\
de Blok, W. J. G., McGaugh, S. S. 1997, MNRAS, 290, 533.\\
de Blok, W. J. G., McGaugh, S. S., Bosma, A., Rubin, V. C. 2001, ApJ, 552L, 23.\\
de Blok, W. J. G., McGaugh, S. S., van der Hulst, J. M. 1996, MNRAS, 283, 18.\\
de Blok, W. J. G., van der Hulst, J. M., Bothun, G. D. 1995, MNRAS, 274, 235.\\
de Blok, W. J. G., van der Hulst, J. M. 1998, A\&A, 336, 49.\\
Dekel, A., Silk, J. 1986, ApJ, 303, 39.\\
Disney, M. J. 1976, Nature, 263, 573.\\
Fermi LAT Collaboration: Abdo, Ackermann, Ajello et al. 2010, ApJ, 712, 147.
Freeman, K. C. 1970, ApJ, 160, 767.\\
Friedli, D., Benz, W. 1993, A\&A, 268, 65.\\
Galaz, G., Herrera-Camus, R., Garcia-Lambas, D., Padilla, N. 2011, ApJ, 728, 74.\\
Galaz, G., Dalcanton, J. J., Infante, L., Treister, E. 2002, AJ, 124, 1360.\\
Geller, M. J., Diaferio, A., Kurtz, M. J., Dell’Antonio, I. P. et al. 2012, AJ, 143, 102.\\
Gerritsen, J. P. E., de Blok, W. J. G. 1999, A\&A, 342, 655.\\
Greene, J. E., Ho, L. C. 2007, ApJ, 670, 92.\\
Grenier, I. A., Casandjian, J.-M., Terrier, R. 2005, Sci, 307, 1292.\\
Gültekin, K., Cackett, E. M., Miller, J. M., Di Matteo, T., Markoff et al. 2009, ApJ, 706, 404.\\
Heckman, T. M., Kauffmann, G., Brinchmann, J., Charlot, S. et al. 2004, ApJ, 613, 109.\\
Hoffman, Y., Silk, J., Wyse, R. F. G. 1992, ApJ, 388L, 13.\\
Impey, C., Bothun, G. 1997, ARA\&A, 35, 267.\\
Jimenez, R., Padoan, P., Matteucci, F., Heavens, A. F. 1998, MNRAS, 299, 123.\\
Kormendy, J., Kennicutt, R. C. Jr. 2004, ARA\&A, 42, 603.\\
Kormendy, J., Freeman, K. C. 2004, IAUS, 220, 377.\\
Kormendy, J. 1990, ASPC, 10, 33.\\
Kuzio de Naray, R., McGaugh, S. S., de Blok, W. J. G. 2008, ApJ, 676, 920.\\
Matthews, L. D., Gallagher, J. S. III, Krist, J. E. et al. 1999, AJ, 118, 208.\\
Matthews, L. D., Gao, Yu 2001, ApJ, 549L, 191.\\
Matthews, L. D., Gao, Y., Uson, J. M., Combes, F. 2005, AJ, 129, 1849.\\
Matthews, L. D., van Driel, W., Monnier-Ragaigne, D. 2001, A\&A, 365, 1.\\
Matthews, L. D., Wood, K. 2003, ApJ, 593, 721.\\
McGaugh, S. S., Bothun, G. D. 1994, AJ, 107, 530.\\
McGaugh, S. S., Rubin, V. C., de Blok, W. J. G. 2001, AJ, 122, 2381.\\
McGaugh, S. S., Schombert, J. M., Bothun, G. D., de Blok, W. J. G. 2000, ApJ, 533L, 99.\\
Mayer, L., Wadsley, J. 2004, MNRAS, 347, 277.\\
McGaugh, S. S., de Blok, W. J. G. 1998, ApJ, 499, 66.\\
McGaugh, S. S., Schombert, J. M., Bothun, G. D. 1995, AJ, 109, 2019.\\
Mihos, J. C., McGaugh, S. S., de Blok, W. J. G. 1997, ApJ, 477L, 79.\\
Mishra, A., Das, M., Kantharia, N. G., Srivastava, D. C. 2013, MNRAS, in preparation.\\
Morelli, L., Corsini, E. M., Pizzella, A., Dalla Bont, E., Coccato, L.,Mndez-Abreu, J., Cesetti,
M. 2012, MNRAS, 423, 962.\\
Naik, S., Das, M., Jain, C., Paul, B. 2010, MNRAS, 404, 2056.\\
Navarro, J. F., Frenk, C. S., White, S. D. M. 1996, ApJ, 462, 563.\\
Norman, C. A., Sellwood, J. A., Hasan, H. 1996, ApJ, 462, 114.\\
O’Neil, K., Bothun, G. D., Schombert, J. 2000, AJ, 119, 136.\\
O’Neil, K., Bothun, G., van Driel, W., Monnier Ragaigne, D. 2004, A\&A, 428, 823.\\
O’Neil, K., Oey, M. S., Bothun, G. 2007, AJ, 134, 547.\\
O’Neil, K., Schinnerer, E. 2003, ApJ, 588L, 81.\\
O’Neil, K., Schinnerer, E., Hofner, P. 2003, ApJ, 588, 230.\\
Ostriker, J. P., Peebles, P. J. E. 1973, ApJ, 186, 467.\\
Pickering, T. E., van Gorkom, J. H., Impey, C. D., Quillen, A. C. 1999, AJ, 118, 765.\\
Pickering, T. E., Impey, C. D., van Gorkom, J. H., Bothun, G. D. 1997, AJ, 114, 1858.\\
Planck Collaboration: Ade, P. A. R., Aghanim, N., Arnaud, M., Ashdown, M. et al. 2011,
A\&A, 536, 19.\\
Pustilnik, S. A., Martin, J.-M., Tepliakova, A. L., Kniazev, A. Y. 2011, MNRAS, 417, 1335.\\
Ramya, S., Prabhu, T. P., Das, M. 2011, ApJ, 728, 124.\\
Rees, M. J. 1984, ARA\&A, 22, 471.\\
Reichard, T. A., Heckman, T. M., Rudnick, G., Brinchmann, J. et al. 2008, ApJ, 677, 186.\\
Rosenbaum, S. D., Krusch, E., Bomans, D. J., Dettmar, R.-J. 2009, A\&A, 504, 807.\\
Sabatini, S., Davies, J., Scaramella, R., Smith, R. et al. 2003, MNRAS, 341, 981.\\
Schombert, J. 1998, AJ, 116, 1650.\\
Schombert, J. M., Bothun, G. D., Impey, C. D., Mundy, L. G. 1990, AJ, 100, 1523.\\
Schombert, J. M., Bothun, G. D., Schneider, S. E., McGaugh, S. S. 1992, AJ, 103, 1107.\\
Schombert, J., Maciel, T., McGaugh, S. 2011, AdAst, 12, 12.\\
Seth, A., Agüeros, M., Lee, D., Basu-Zych, A. 2008, ApJ, 678, 116.\\
Silk, J., Rees, M. J. 1998, A\&A, 331, L1.\\
Sprayberry, D., Impey, C. D., Bothun, G. D., Irwin, M. J. 1995, AJ, 109, 558.\\
Sprayberry, D., Impey, C. D., Irwin, M. J., McMahon, R. G., Bothun, G. D. 1993, ApJ, 417,
114.\\
Subramanium, S., Das, M., Sivarani, T., Ramya, S. et al. 2013, MNRAS, in preparation.\\
Swaters, R. A., Madore, B. F., Trewhella, M. 2000, ApJ, 531L, 107.\\
Trachternach, C., Bomans, D. J., Haberzettl, L., Dettmar, R.-J. 2006, A\&A, 458, 341.\\
van Moorsel, G. A. 1982, A\&A, 107, 66.\\
van der Hulst, J. M., Skillman, E. D., Smith, T. R., Bothun, G. D. et al. 1993, AJ, 106, 548.\\
Wyder, T. K., Treyer, M. A. 2011, ASPC, 440, 235.\\

\newpage

\begin{table}[h]
\vspace{-3mm}
\centering
\caption{\bf Examples of GLSB Galaxies}
\begin{tabular}{lcccccccc}
\hline\hline
Galaxy           & Distance & ${\mu_{0}}^a$ & Bulge/Disk & M(HI) & Reference \\
Name             &   (Mpc)  & and Waveband  & lum. Ratio & $10^{10}~M_{\odot}$ &      \\
\hline
Malin~1          & 366      & 26.0, R~band & 0.40  & 6.8  & P97 \\ 
Malin~2 (F568-6) & 201      & 23.4, B~band & 0.84  & 3.6 & P97 \\
UGC~6614         & 93.2     & 24.3, B~band & 1.35  & 2.5 & P97 \\
NGC~7589         & 120      & 23.3, R~band & 1.78  & 0.9  & P97 \\
UGC~2936         & 51.2     & 24.0, B band & 0.83  & 0.3 & S95 \\
1226+0105        & 349      & 23.3, B band & 0.51  & 1.8 & S95 \\
2327-0244 (UM~163) & 136    & 23.2, B band & 0.59  & 0.1  & B01 \\
\hline
\end{tabular}
a: extrapolated disk surface brightness.\\
P97: Pickering et al. 1997.\\
S95: Sprayberry et al. 1995.\\
Burkholder et al. 2001
\end{table}

\vspace{6mm}

\begin{table}[h]
\vspace{-3mm}
\centering
\caption{\bf List of Galaxies, types and black holes masses}
\begin{tabular}{lccccccc}
\hline\hline
Galaxy           & Redshift & Hubble Class      & Black Hole Mass       & Reference \\
                 &          & and Nuclear Type  & (10$^{6}$ solar mass) &           \\
\hline
Malin~2 (F568-6) & 0.046    &  Sd/p, Sy~1       & 0.3    & R2011 \\ 
UGC~6614         & 0.021    &  (R)SA(r)a, Sy~1  & 3.9    & R2011 \\
UGC~6968         & 0.028    &  S, Sy~1          & 0.5    & R2011 \\ 
UGC~1922         & 0.036    &  S, Sy~1          & 0.4    & R2011 \\
1226+0105        & 0.079    &  Sc, Sy~1         & 12.7   & S2013 \\
2315-0000        & 0.030    &  SAB(rs)a, Sy~1   & 3.8    & S2013 \\ 
Malin~1          & 0.083    &  S, Sy~1          & 0.5    & S2013 \\ 
2MASX~J09593953+0035117 & 0.066 & Irr, Sy~1     & 1.0    & S2013 \\
SDSS~J135943.13-003424.4 & 0.164 & Sc, NLSy1    & 19.9   & S2013 \\ 
\hline
\end{tabular}
R2011 : Ramya et al. 2011.\\
S2013 : Subramanium et al. 2013.
\end{table}

\vspace{3mm}

\begin{figure}[ht]
\centering
\vspace*{7.2cm}
\leavevmode
\includegraphics{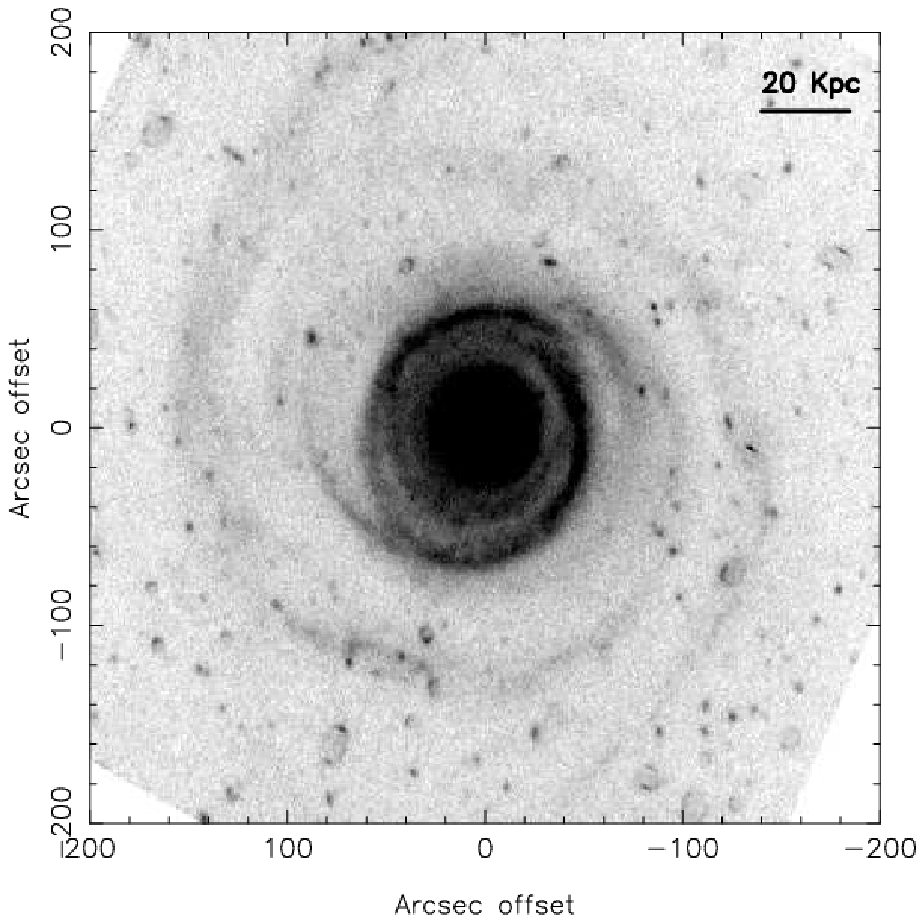}
\includegraphics{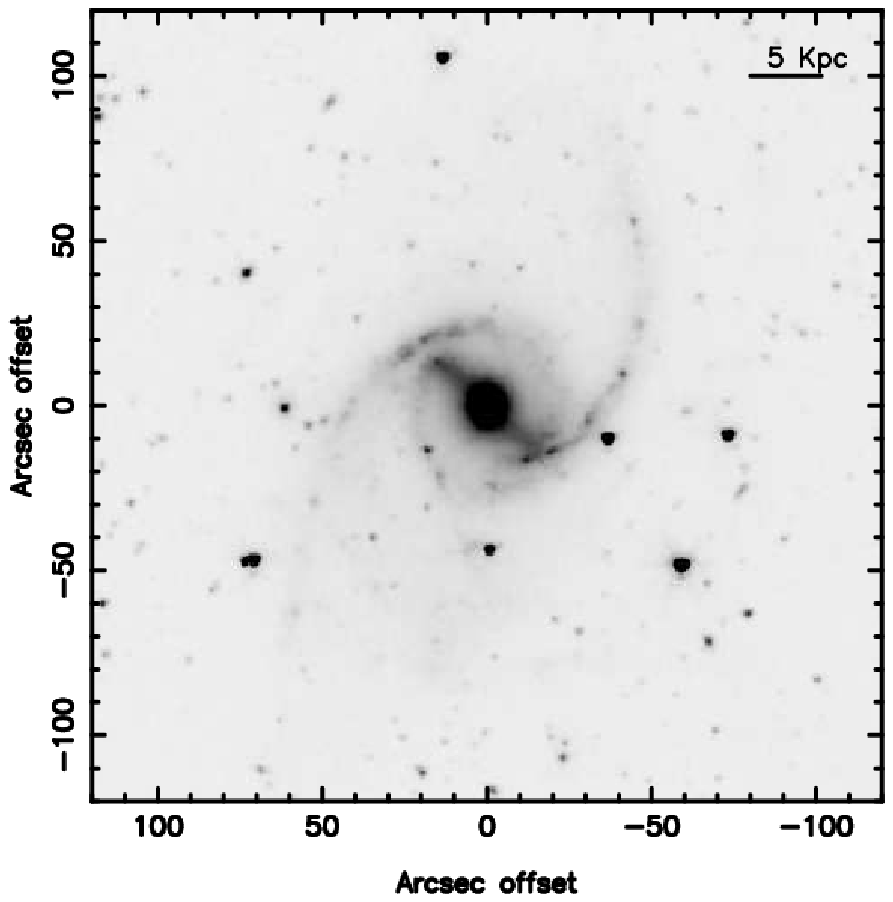}
\vspace{-2.6cm}
\caption{
(a)~Figure on the left shows the R band image of the GLSB galaxy UGC~6614. The bar on the top right corner gives the lengthscale in Kpc.
The galaxy has a very prominent bulge and tightly wound spiral arms that form a ring like structure outside the
bulge. The LSB disk extends well beyond the plot boundaries. 
(b)~Figure on the right is the 3.6 micron IRAC mid-infrared image of the GLSB galaxy NGC~5905. This is one of the few GLSB galaxies that 
is also interacting with a nearby companion galaxy (NGC~5908). The interaction has probably induced the formation of the bar and spiral
arms. }   
\vspace{-4mm}
\end{figure}

\vspace{3mm}

\begin{figure}[ht]
\centering
\vspace*{7.2cm}
\leavevmode
\includegraphics{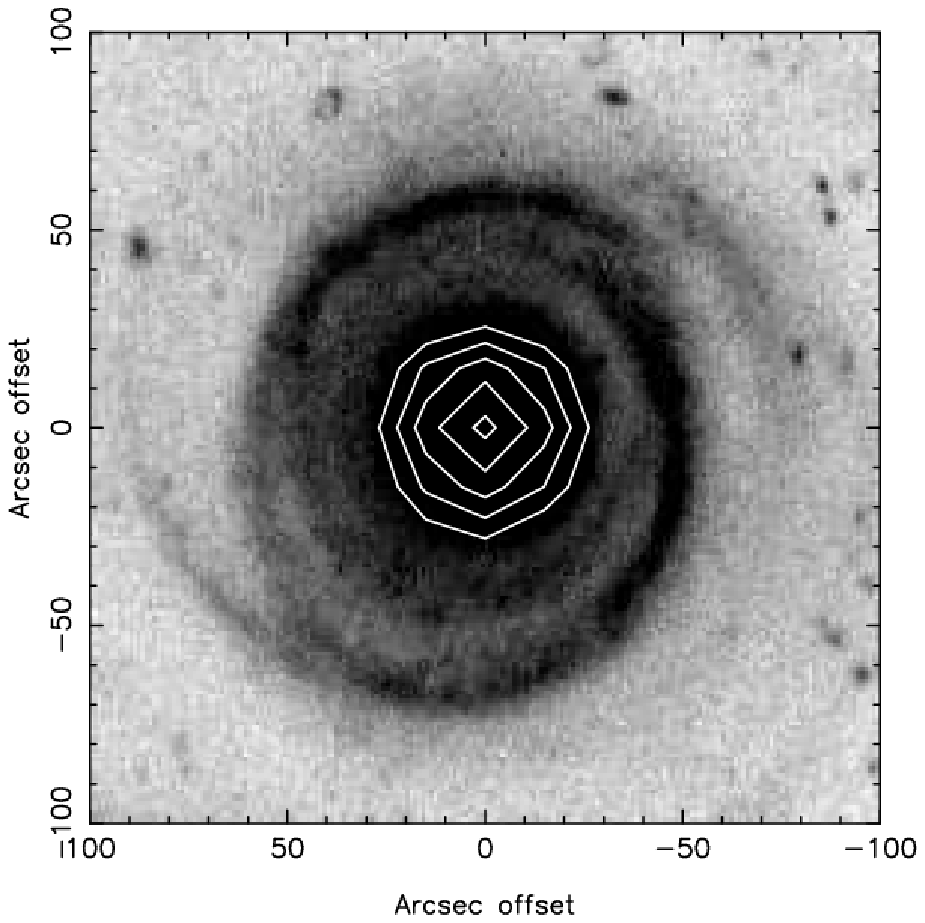}
\includegraphics{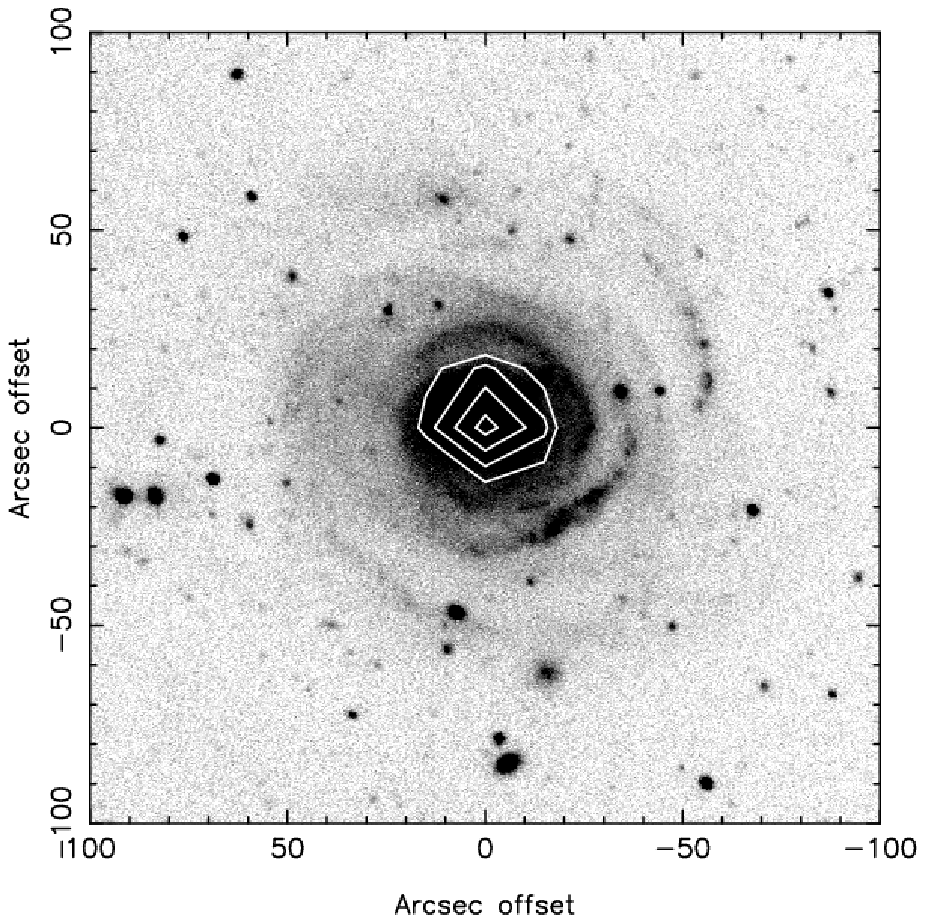}
\vspace{-2.6cm}
\caption{
(a) The contours of VLA 1.4~GHz continuum emission from UGC~6614 overlaid on the R band image of the galaxy. The
figure shows some extended features which could be due to outflows associated with the AGN. This is clearer in the
610~GHz observations (Mishra et al. 2013, in preparation).
(b) The contours of VLA 1.4~GHz continuum emission from Malin~2 overlaid on the 2MASS K band image of the galaxy. 
The compact emission is centered on the galaxy nucleus and does not show extended features at 610~GHz 
(Mishra et al. 2013, in preparation). }
\vspace{-4mm}
\end{figure}

\end{document}